# Uncorrelated binary sequences of lengths $2^a3^b4^c5^d7^e11^f13^g$ based on nested Barker codes and complementary sequences

Patricio G. Donato, Matías N. Hadad and Marcos A. Funes

*Abstract*— **Certain applications require the use of signals that combine both the capability to operate with low signal-to-noise ratios and the ability to support multiple users without interference. In the case where many users have very different signal-to-noise ratios, it is necessary to consider coding schemes that can be used in a multi-user environment but with different noise immunity levels. Traditional detection systems based on the correlation function and coding sequences have significant limitations in satisfying both objectives, since the cross-correlation between coded signals corresponding with different users is linked to the use of the same coded sequences' length. The research topic of binary sequences that have null cross-correlation and different length has not been studied in depth, but it has potential applications in multi-user environments. In this work an algorithm to generate binary sequences completely uncorrelated with certain sets of complementary sequences is presented. The proposed algorithm is based on nested Barker sequences, and it is compared with a previous proposal based on an iterative algorithm. This approach allows to generate more diversity of sequences of different length than the iterative approach, which it makes useful for applications based on binary sequences detection and expand the horizon of many applications. [1]**

## I. INTRODUCTION

BINARY complementary sequences have been used in many different fields of engineering, in particular to improve the Signal-to-noise Ratio (SNR) of heavily attenuated and/or interfered signals. Their applications range from radar/sonar to communications, including transmission channel identification and non-destructive tests [1] [2]. The focus of the research has been concentrated mainly in the application of these sequences in each field and the design of efficient logic algorithms for the generation, processing and post-processing [3] [4]. One useful characteristic of these sequences, in addition to improving SNR, is the possibility of synthesize pairs and sets of sequences that are mutually uncorrelated, which are useful to employ in multi-user systems, where many coded signals must share the same physical channel, minimizing the mutual interference. However, there are practical cases where this capability of identification may be conditioned, as the case when different users are affected by different SNR levels. In those situations, some sequences could not be identified. Classical examples are the Local Positioning Systems (LPS), where a receiver determines its positions through the time calculation from signals emitted by fixed beacons. In those applications it is common to deal with a low SNR and positioning is not possible because the signals of some beacons cannot be detected [5] [6]. Other examples of positioning systems with high noise sensitivity can be found in applications very different, as the mouse 3D of La Cour-Harbo & Stroustrup [7] and Unmanned Aerial Vehicles (UAV) positioning. Furthermore, there are other very different applications where immunity to noise and the use of uncorrelated sequences is useful, such as railway safety systems [8] or cognitive radio networks [9]. In order to avoid the use of sets of complementary sequences of larger length, which leads to longer transmission times, and consequently lower transmission rates for all users, it would be desirable to use some type of coding sequences that maintain a null cross-correlation even though the length of the sequences is different. This topic has not been thoroughly explored, and the main efforts to generate uncorrelated sets of sequences have been concentrated on polyphase sequences [14]. However, binary sequences are simpler than polyphase and the algorithms for generation and processing are more easily implemented, with lower resources requirements. Additionally, binary sequences are easily modulated and adapted to the transmission medium, reducing the requirements for their detection, in comparison with non-binary approaches that involves more complex algorithms and modulation techniques (because it is necessary to compute real and imaginary values). The research interest on binary sequences, being complementary sequences or not, is evidenced in some present works, in which are explored different generation algorithms to synthesize sequences with lengths that are not a

[1]   P.G. Donato (*Instituto de Investigaciones Científicas y Tecnológicas en Electrónica (ICYTE), CONICET-UNMDP, Mar del Plata, Argentina*).
E-mail: donatopg@fi.mdp.edu.ar

M.N. Hadad (*Instituto de Investigaciones Científicas y Tecnológicas en Electrónica (ICYTE), CONICET-UNMDP, Mar del Plata, Argentina*).
E-mail: mhadad@fi.mdp.edu.ar

M.A. Funes (*Instituto de Investigaciones Científicas y Tecnológicas en Electrónica (ICYTE), CONICET-UNMDP, Mar del Plata, Argentina*).
E-mail: mfunes@fi.mdp.edu.ar

power of two [10] [11]. In this work a simple algorithm to obtain uncorrelated binary complementary sequences of different length using a Barker approach is proposed. The main goal is the generation of uncorrelated sets of sequences able to use in multi-user systems. This proposal is not linked to a particular application, but is a mathematical approach useful in different fields as previously cited [5] [6] [8].

## II. STATE OF THE ART ON COMPLEMENTARY SEQUENCES AND UNCORRELATED SEQUENCES OF DIFFERENT LENGTH

In this section is presented a brief summary of concepts and previous contributions that can be found in the specific bibliography, which are necessary to introduce the proposal. First the complementary sequences are defined, which constitute the basis of this study, and subsequently the previous contributions to the problem of uncorrelated sequences of different lengths are discussed.

### II.a. Basic definitions of Complementary Sequences

Complementary sequences [12] can be defined as sets of $M=2^m$ binary sequences of length $L=M^N$, denoted with the matrix $\mathbf{S}_{M,L}$.

$$\mathbf{S}_{M,L} = \begin{bmatrix} S_1 \\ S_2 \\ \cdots \\ S_M \end{bmatrix} \Rightarrow S_j = \begin{bmatrix} S_{j,1} & S_{j,2} & \cdots & S_{j,L} \end{bmatrix} \tag{1}$$

Although there are complementary sequences of lengths that are powers of 10 and 26, they are limited to sets of only two sequences ($m=1$) and they will not be considered for this work. Complementary sequences of any length $L$ are characterized by two main mathematical properties:

- The sum of the autocorrelations of all the sequences is a Kronecker delta of amplitude $M \cdot L$ [13]:

$$SAC[k] = \sum_{j=1}^{M} \sum_{i=1}^{L} S_j[k] \cdot S_j[i+k] = M \cdot L \tag{2}$$

- Given a set of complementary sequences, there are $M$-1 sets with the same length that are mutually uncorrelated. For example, considering two uncorrelated sets $\mathbf{S}^{\alpha}_{M,L}$ and $\mathbf{S}^{\beta}_{M,L}$ [13]:

$$SCC[k] = \sum_{j=1}^{M} \sum_{i=1}^{L} S_j^{\alpha}[k] \cdot S_j^{\beta}[i+k] = 0 \tag{3}$$

In practical applications, a more accurate detection is achieved by increasing the length of all sequences in the set ($L$) or by increasing the number of sequences in the set ($M$), which substantially increases the transmission period and the processing time. Alternatively, some authors have tried to generate sequences with different lengths to address this problem, using approaches with dissimilar complexity.

### II.b. Previous contributions on the topic of uncorrelated sequences of different length

There are few examples in the specific bibliography about the synthetization of uncorrelated sequences of different length. The work of Raja Durai *et al.* [14] proposed an algorithm that is restricted to sets of $4^m$ (being $m$ a natural number) complementary sequences composed by quaternary elements, that is to say, four different types of elements (-1, +1, -$j$, +$j$). Algorithms of lower complexity have been obtained with the concatenation of uncorrelated sets of complementary sequences, denominated Complementary-Derived Orthogonal Sequences, CDOS [15], that allow the construction of sets of binary sequences of length $L'=2^p \cdot L$. Basically, the algorithm uses an iterative concatenation process alternating signs, which doubles the length in each iteration $p$, as it is shown in the following example:

$$p = 0 \rightarrow \mathbf{S'}_{M,L} = \mathbf{S}_{M,L}$$

In the first iteration ($p = 1$), the CDOS doubles the length of the original sequence:

$$p = 1 \rightarrow \mathbf{S'}_{M,2L} = [+\mathbf{S}_{M,L} \ -\mathbf{S}_{M,L}]_{M,2L}$$

In the second iteration ($p = 2$), the CDOS doubles the length of the previous iteration, that is to say, quadruples the original sequence:

$$p = 2 \rightarrow \mathbf{S'}_{M,4L} = [+[+\mathbf{S}_{M,L} \ -\mathbf{S}_{M,L}] \ -[+\mathbf{S}_{M,L} \ -\mathbf{S}_{M,L}]]_{M,4L}$$

In the Q-th iteration ($p = Q$), the CDOS multiplies by $2^Q$ the length of the original sequence:

$$p = Q \rightarrow \mathbf{S'}_{M,2^{Q}L} = [+\mathbf{S'}_{M,2^{(Q-1)}L} \ -\mathbf{S'}_{M,2^{(Q-1)}L}]_{M,2^{Q}L}$$

The extended sequence, $\mathbf{S'}_{ML}$, is uncorrelated with all the other complementary sets of sequences of length $L$, that is, it satisfies equation (3). This approach can be applied to sets of any number of sequences or length, but the concatenated sequences does not match with the correlation conditions that define a set of complementary sequences (see equation 2), namely the sum of the autocorrelation functions of the new set is not a single Kronecker delta, but contains sidelobes with different amplitudes and polarity.

There are not any other algorithms proposed to deal with the problem of multi-user systems with different SNRs that are based on binary sequences. For this reason, in that follows, the comparison will be made with the CDOS approach.

### III. Proposed algorithm based on Barker codes and Complementary Sequences

To extend the number of lengths that are potentially synthesizable, it is necessary to use a logical algorithm not based on powers of 2 or 4, as in the case of the previous contributions [14] [15]. The use of power of 2 or 4 generates few different sequences' lengths, which can be avoided using a combination of different powers. The proposal is to use the well-known Barker sequences to determine the signs (+ or -) of the sequences in the concatenation process, assuming that, as with CDOS, the complementary correlation property (equation 2) is not required. Barker sequences are characterized by a mainlobe of amplitude equal to the length of the sequence and sidelobes of unitary amplitude. The main constraint of these sequences is the reduced number of lengths available, which are limited to sequences of 2, 3, 4, 5, 7, 11 and 13 elements (see Table 1). Some authors have demonstrated the absence of binary Barker sequences of length $L_B > 13$ [16].

Table 1: Binary Barker sequences

| Length | Barker sequences |
|--------|------------------|
| 2 | [+1 +1] or [+1 -1] |
| 3 | [+1 +1 -1] |
| 4 | [+1 +1 +1 -1] or [+1 +1 -1 +1] |
| 5 | [+1 +1 +1 -1 +1] |
| 7 | [+1 +1 +1 -1 -1 +1 -1] |
| 11 | [+1 +1 +1 -1 -1 -1 +1 -1 -1 +1 -1] |
| 13 | [+1 +1 +1 +1 +1 -1 -1 +1 +1 -1 +1 -1 +1] |

The extended set of sequences can be generated considering that each element of a Barker sequence $B$ is a set of complementary sequences of length $L$, as follows:

$$\mathbf{S'}_{M,L'} = B(z^L) \cdot \mathbf{S}_{M,L} \tag{4}$$

where the matrix $\mathbf{S}_{M,L}$ is a set of $M = 2^m$ complementary sequences of length $L = M^N$ binary elements and $B(z)$ is a Barker sequence considered as a sequence in the discrete domain $z$. For example, for a Barker sequence of length $L_B = 4$ (see Table 1 for the Barker sequences):

$$B = [+1 +1 +1 -1]$$
$$B(z) = 1 + z^{-1} + z^{-2} - z^{-3}$$
$$\mathbf{S'}_{M,L'} = \left(1 + z^{-L} + z^{-2L} - z^{-3L}\right) \mathbf{S}_{M,L}$$
$$\mathbf{S'}_{M,L'} = \left(\mathbf{S}_{M,L} + \mathbf{S}_{M,L} \cdot z^{-L} + \mathbf{S}_{M,L} \cdot z^{-2L} - \mathbf{S}_{M,L} \cdot z^{-3L}\right) \tag{5}$$

The sum of the autocorrelation functions of the extended set of sequences $\mathbf{S'}_{M,L'}$ (for the particular case of $L' = 4 \cdot L$ and a generic $M$) is shown in Fig. 1. In order to compare with the approach of CDOS, an extended set of $M$ sequences of length $4 \cdot L$ generated with the iterative process is shown. Note that the amplitude of the mainlobe is equal for both extended sequences, but the sidelobes are different, both in amplitude and polarity. Considering only the absolute value, it is observed, in the case of the Barker sequences, that the ratio between the mainlobe and the sidelobes is higher than for the case of the sequence generated with the iterative process.

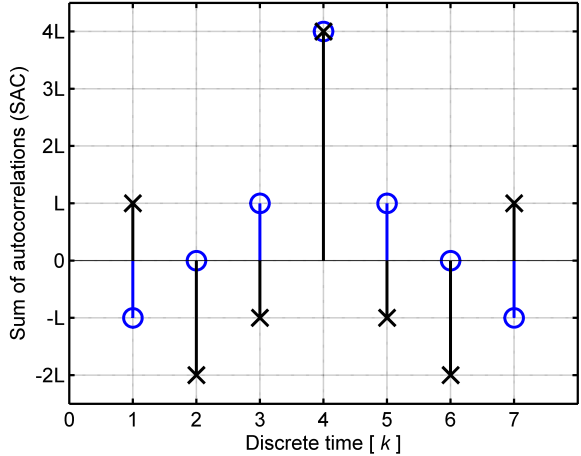

Fig. 1 Sum of the autocorrelation functions of the set of sequences extended with a Barker approach (blue circle markers) compared with the corresponding to a CDOS of the same length (black cross markers). Note that mainlobe remains equal in both cases, but the absolute value of sidelobes amplitude is lower with Barker than CDOS approach.

However, the proposal seems to be insufficient, because the possible length increments are limited to multiples of the binary Barker sequences lengths. To obtain more diversity of lengths it is necessary to combine different Barker sequences, using a nesting arrangement where each element of a Barker sequence is another Barker sequence of equal or different length. The combination of Barker codes of different lengths was considered by J.M. Howell [17], who limited his work to the study from the point of view of the Doppler Effect sensitivity. However, its application as a technique for improving SNR and as an algorithm for sequences generation with null cross-correlation was not analyzed. Another contribution in that sense was presented by Z. Fan *et al.* [18], using a convolution of Barker and Golay sequences. In this case, the author demonstrates the signal to noise ratio improvement for ultrasonic applications, but the use in multi-user schemes with different SNR is not considered and the properties of orthogonality and uncorrelation among different sets of sequences are not discussed. The proposal of the present work is mainly concentrated in these last topics.

To better understand the nesting arrangement proposed here, consider the next example. Supposing a Barker sequence of length $L_1$=4 and other of length $L_2$=5, it is possible to obtain a nested sequence of length $L_3$=$L_1 \cdot L_2$=20, as follows:

Barker sequence of length 4 $\rightarrow \begin{bmatrix} +1 & +1 & +1 & -1 \end{bmatrix}$

Barker sequence of length 5 $\rightarrow \begin{bmatrix} +1 & +1 & +1 & -1 & +1 \end{bmatrix}$

Nested Barker sequence of length 20 = (4·5)
$\begin{bmatrix} +1 +1 +1 -1 +1 +1 +1 -1 +1 +1 +1 -1 -1 -1 -1 +1 +1 -1 \end{bmatrix}$

Using this arrangement, extended to the case of more than two nested sequences, it is possible to generate new sequences of lengths that are multiples of all the lengths of Barker sequences:

$$L' = 2^a \, 3^b \, 4^c \, 5^d \, 7^e \, 11^f \, 13^g \qquad (6)$$

being the exponents $a$, $b$, $c$, $d$, $e$, $f$ and $g$ natural numbers. Using different combinations of exponents it is possible to obtain different lengths. For example, considering the range from 1 to 128, using nested Barker codes it is possible to synthesize 73 different extended sequences, that is to say, 57% of the lengths comprehended in this range. The sequences that can not be synthesized are those corresponding to lengths equal to prime numbers greater than 13 or composite numbers that are only divisible by prime numbers greater to 13. On the contrary, using the CDOS approach [15] or any approach based on lengths proportional to powers of two (any algorithm based on Hadamard matrices), only 7 different lengths (2, 4, 8, 16, 32, 64 and 128) can be synthesized, which represents the 5,4% of the lengths available between 1 and 128.

## IV. SIDELOBE TO MAINLOBE RATIO (SMR)

To evaluate the advantages of the proposal based on nested Barker sequences to generate uncorrelated complementary sequences of different length, besides considering the available lengths, it is important to calculate the sidelobe to mainlobe ratio (SMR). This ratio is related with the ability to discriminate a mainlobe from a sidelobe and, therefore, the mainlobe from the noise. An SMR close to 0 means that the sidelobes are not relevant with regard to the amplitude of the mainlobe, making the mainlobe clearly distinguishable from the noise. On the contrary, an SMR close to 1, means that sidelobes and mainlobe have similar amplitudes. In this case, the mainlobe is hardly distinguishable from the surrounding sidelobes or the noise.

Figure 2 shows the evolution of the SMR as a function of the length, considering the maximum in absolute value of the sidelobes. While using the method of concatenation with sign inversions [15] the ratio is always ½, for the case of Barker ordering logic this is the maximum ratio. In this case, the maximum sidelobe obtained in the sum of the autocorrelation functions is, in the worst case, a half of the mainlobe, like the original CDOS, but with the advantage that greater flexibility in length is obtained. The SMR values spread is caused by the fact that when Barker sequences are sequentially nested, the shortest sequence determine the total SMR. In the case of even lengths, a sequence of length 2 is usually used in the process (except in the case of multiples of 4), which conditions the SMR to a value of ½. The best SMR is obtained for the longest single Barker sequence, namely, for the 13-element Barker, where an SMR=1/13 is obtained, or any iterative nesting of length 13 ($13^2$, $13^3$, etc.).

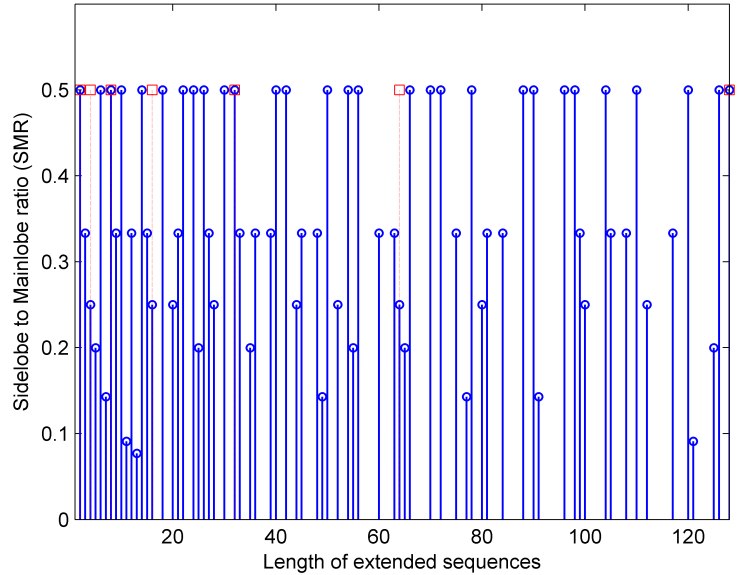

Fig. 2 Sidelobe to mainlobe ratio (SMR) for extended sequences generated with the Barker approach (blue circles) and the CDOS approach (red squares). Note that the Barker nested approach allows a more diversity of lengths, with lower (better) or equal SMR than CDOS.

To better understand the advantages of the length flexibility provided by Barker nested sequences, let consider a case of a local positioning system with four beacons. To allow the univocal identification of each individual beacon, each signal is encoded with a particular complementary sequences set that is uncorrelated with all the other beacons. For simplicity, let consider sets of $M$=4 complementary sequences of length $L$=4 that satisfy the equations (2) and (3).

Supposing that one of the beacons (for example, the beacon β) is affected by a Signal-to-Noise ratio (SNR) 10 times lower than beacon α, and other beacon (for example γ) is affected by an SNR 6 times lower than beacon α. In this situation, using CDOS it is necessary to synthesize uncorrelated sequences of lengths 16·$L$ and 8·$L$, respectively, in order to increment the SMR a magnitude equal or higher than the SNR decrement. The consequence of this length increment is a reduction in the sequences transmission rate, because it is necessary more time to transmit a coded signal. On the other hand, the use of the Barker approach allows to synthesize uncorrelated sequences of lengths 10·$L$ and 6·$L$, respectively, which increment the SMR a magnitude equal to the SNR reduction, optimizing the transmission rate with regard to the CDOS sequences (Fig. 3). In both cases, CDOS and Barker, the SMR is the same, but the lengths of the uncorrelated sequences obtained are 25% and 37,5% shorter in the case of Barker. Moreover, with the Barker approach, the length can be increased a little bit, synthesizing sequences of length 7 and 11, which reduces the SMR notoriously, maintaining lower length values than those of the CDOS approach (Fig. 4).

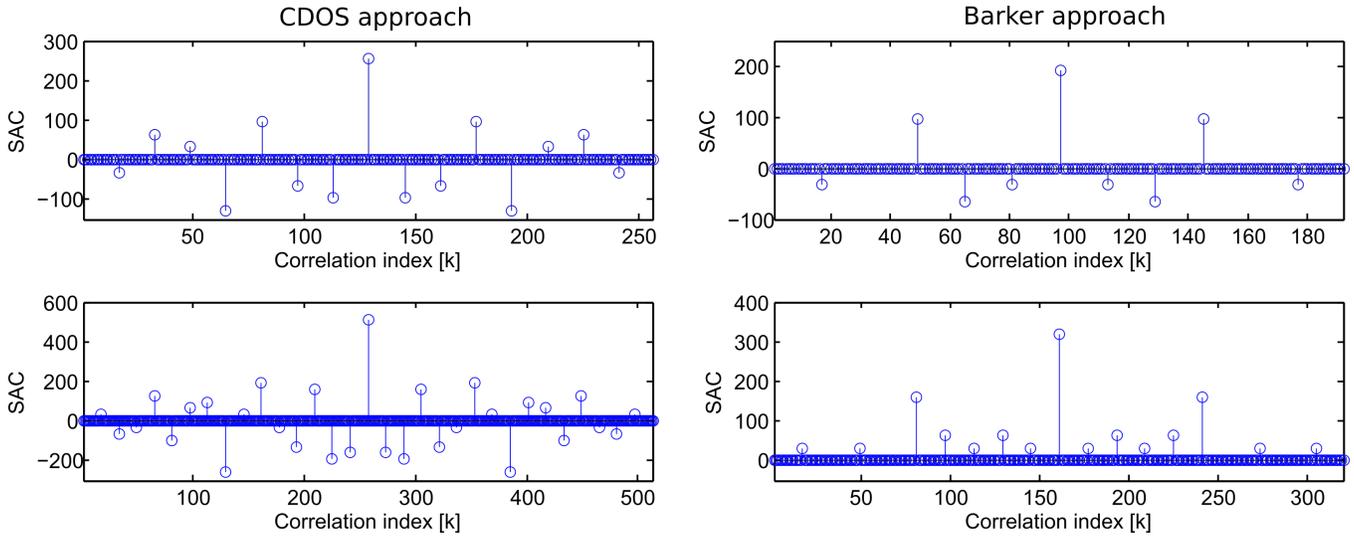

Fig. 3 Sum of the autocorrelation functions of different lengths of CDOS and Barker nested sequences. In all the cases the SMR is equal to ½ but in the case of CDOS approach (left) the sequences are larger than the obtained in the Barker approach (right).

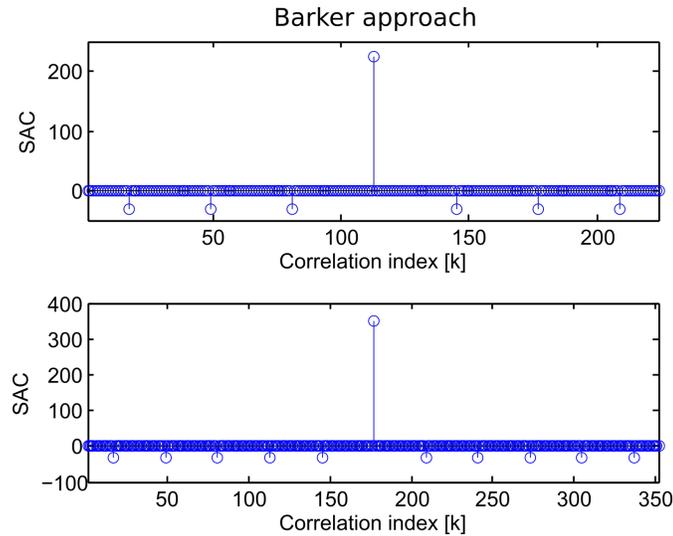

Fig. 4 Sum of the autocorrelation functions of Barker nested sequences of lengths 7·L (top) and 11·L (bottom). In both cases the SMR is improved regarding to the case of Fig. 3, obtaining ratios of 1/7 and 1/11 respectively. In this case, the obtained sequences are shorter than the CDOS, but the SMR is lower, which improves the noise immunity.

An alternative metric to evaluate the advantage of the proposal instead of CDOS approach is the ratio between the amplitude of the mainlobe and the sum of all the sidelobes. This is a useful parameter to evaluate the contrast between the mainlobe and sidelobes. In some applications, i.e. ultrasonic sensor signal processing, the amount of sidelobe contribution to the total correlation is a measure of the strength of the scatters in the detection field [19]. In that sense, it is desirable to remove or at least minimize the contribution of the sidelobes of the used sequences to the correlation. The sum of the absolute value of all the sidelobes can be understood as a measure of this contribution, and hence, a parameter that weights the efficiency of the correlation of the sequences. Calculating this sum, normalized to the mainlobe amplitude, shows that the nested Barker approach has a better ratio than CDOS. This behavior may be explained through the sidelobes distribution of both types of sequences. In the case of Barker sequences the sidelobes always are of unitary value and, in the case of the nesting process, the amplitude of the sidelobes is increased, but not in the same proportion as with CDOS. Since the ordering of CDOS is based in a power of two concatenation rule that, in absolute value, generates more sidelobes: ½ of the mainlobe, ¼ of mainlobe, 1/8, etc. In Fig. 5 some of the cases where can be applied both the Barker and CDOS approaches are shown (lengths 2, 4, 8, 16, 32, 64 and 128). In the top of the figure is shown the sum of all the sidelobes both for CDOS (black line) as for Barker (blue line). As can be seen, the contribution of the sidelobes to the complete correlation signal is equal or bigger for CDOS than Barker. In the bottom is shown the sum of the absolute values of all the sidelobes. The sum of autocorrelations of CDOS is higher than the corresponding to the Barker nested sequences. In both cases is evidenced that the contribution of nested Barker sequences autocorrelation sidelobes is lower than the corresponding to CDOS, which represent a positive feature of this proposal with regard to the approaches based on rows or columns of Hadamard matrices.

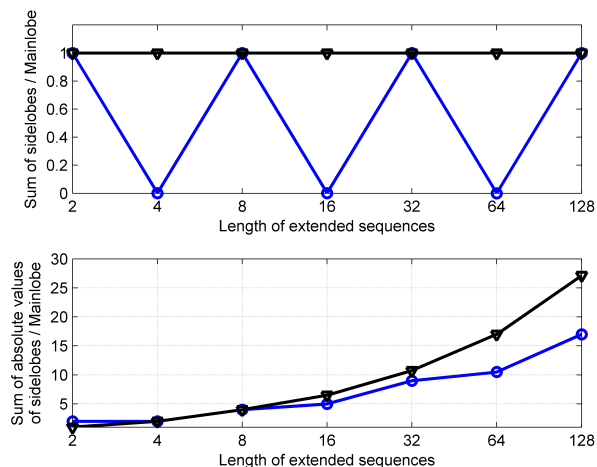

Fig. 5 Ratio of the sum of all the sidelobes of the sum of autocorrelations of CDOS (black line) and Barker nested sequences (blue line) with regard to the mainlobe. In the top plot is shown the signed sum and the bottom corresponds to the sum of the absolute values of the sidelobes.

## V. CONCLUSION

The synthesis of uncorrelated binary sequences of different length is a topic where there are not definitive solutions. In this work it is proposed a simple algorithm that can be used with sets of binary complementary sequences to generate a new uncorrelated set of different length. Although this algorithm is not linked to a specific application, it may be useful for LPS or UAV positioning, among other applications. This algorithm is simpler than previous contributions based on quaternary sequences that can be found in the specific bibliography. With regard to the approach of CDOS, the computational complexity of this algorithm is similar to those sequences, but with better performance. The algorithm allows to generate a wider diversity of lengths, and it has a better ratio between mainlobe and sidelobes after the correlation processing. The nesting approach proposed here can be extended to non-binary complementary sequences, without the loss of any property.

### ACKNOWLEDGEMENT

This work was supported by the Consejo Nacional de Investigaciones Científicas y Técnicas (CONICET), Argentina, and by the Universidad Nacional de Mar del Plata (UNMDP), Argentina.